\documentclass[11pt]{article}
\usepackage[margin=1in]{geometry}
\usepackage{graphicx}
\usepackage{subfigure}
\usepackage{makecell}
\usepackage{amsmath}
\usepackage{amssymb}
\usepackage{comment}
\usepackage{longtable}
\usepackage[hyperfootnotes=true]{hyperref}
\usepackage{float}
\hypersetup{
 colorlinks=true,
 citecolor=blue,
 linkcolor=blue,
 urlcolor=blue}

\def\beq{\begin{equation}}
\def\eeq#1{\label{#1}\end{equation}}
\def\eeqn{\end{equation}}

\def\beqa{\begin{eqnarray}}
\def\eeqa#1{\label{#1}\end{eqnarray}}
\def\eeqan{\end{eqnarray}}

\let\bar=\overbar

\def\Dslash{\not{\hbox{\kern-4pt $D$}}}
\def\dslash{\not{\hbox{\kern-2pt $\del$}}}

\def\msb{{\bar{\ssstyle M \kern -1pt S}}}

\def\Title#1{\begin{center} {\Large {\bf #1} } \end{center}}
\def\Author#1{\begin{center} {\normalsize {\sc #1} } \end{center}}
\def\Institution#1{\begin{center} {\normalsize {\it #1} } \end{center}}
\def\Abstract#1{\noindent {\normalsize {\bf Abstract:} {\normalfont #1}}}
\def\Conference{\vspace{4mm}\begin{raggedright} {\normalsize {\it Talk presented at the 2019 Meeting of the Division of Particles and Fields of the American Physical Society (DPF2019), July 29--August 2, 2019, Northeastern University, Boston, C1907293.} } \end{raggedright}\vspace{4mm}}

\begin{document}

%%%%%%%%%%%%%%%%%%%%%%%%%%%%%%%%%%%%%%%%%%%%%%%%%%%%%%%%%%%%%%%%%%%%%%%%%%%
%
% TITLE, AUTHOR, INSTITUTION, ABSTRACT ==> UPDATE
% 
%%%%%%%%%%%%%%%%%%%%%%%%%%%%%%%%%%%%%%%%%%%%%%%%%%%%%%%%%%%%%%%%%%%%%%%%%%%

\Title{Extending RECAST For Truth-Level Reinterpretations}

\Author{Alex Schuy\textsuperscript{1}, Lukas Heinrich\textsuperscript{2}, Kyle Cranmer\textsuperscript{3} and Shih-Chieh Hsu\textsuperscript{1}}

\Institution{\textsuperscript{1}University of Washington, Seattle, USA \\ \textsuperscript{2} European Laboratory for Particle Physics, CERN \\ \textsuperscript{3} New York University, New York, USA }

\Abstract{RECAST is an analysis reinterpretation framework; since analyses are often sensitive to a range of models, RECAST can be used to constrain the plethora of theoretical models without the significant investment required for a new analysis. However, experiment-specific full simulation is still computationally expensive. Thus, to facilitate rapid exploration, RECAST has been extended to truth-level reinterpretations, interfacing with existing systems such as RIVET.
}

\Conference

%%%%%%%%%%%%%%%%%%%%%%%%%%%%%%%%%%%%%%%%%%%%%%%%%%%%%%%%%%%%%%%%%%%%%%%%%%%
%
% MAIN TEXT ==> UPDATE
% 
%%%%%%%%%%%%%%%%%%%%%%%%%%%%%%%%%%%%%%%%%%%%%%%%%%%%%%%%%%%%%%%%%%%%%%%%%%%

\section{Introduction}

At the LHC, searches for extensions of the standard model (SM) of particle physics, so-called 'beyond the standard model' (BSM) theories, are undertaken using a complex workflow incorporating event simulation, selection, and statistical analysis. In general, each search requires a new set of event selections, in order to focus on a region of kinematic phase space in which there is a significant, detectable difference between the predictions of the BSM and SM theories. Development of these event selections, along with a corresponding method of rigorous statistical analysis, can take years to complete. In an effort to reduce this cost, work has been done in recent years to develop the capability to preserve and reinterpret existing analyses in a software framework known as RECAST~\cite{recast}.

\section{Collaboration-specific Reinterpretation and Preservation}

Work on RECAST originated in the ATLAS collaboration at CERN. RECAST was designed to solve two concerns that ATLAS had:

\begin{enumerate}
    \item The number of BSM models that need to be considered is growing faster than the available personnel resources for analysis design.
    \item Use of the software related to an analysis often requires expert knowledge only present on the original team, which has contributed to a reproducibility crisis.
\end{enumerate}

For the first item, it was realized that analyses often utilize signatures that make them sensitive to a wide range of models, more than are initially explored. Thus, if search workflows are written in a modular fashion that is easily usable by future collaborators, then progress can be made on more searches more quickly by reusing existing analyses. Fortunately, eliminating the need for expert knowledge by making standardized modular search workflows also solves the reproducibility problem, so we see that we can solve both issues with the same system.\\

\begin{figure}
    \centering
    \subfigure[]{
    \includegraphics[width=\textwidth]{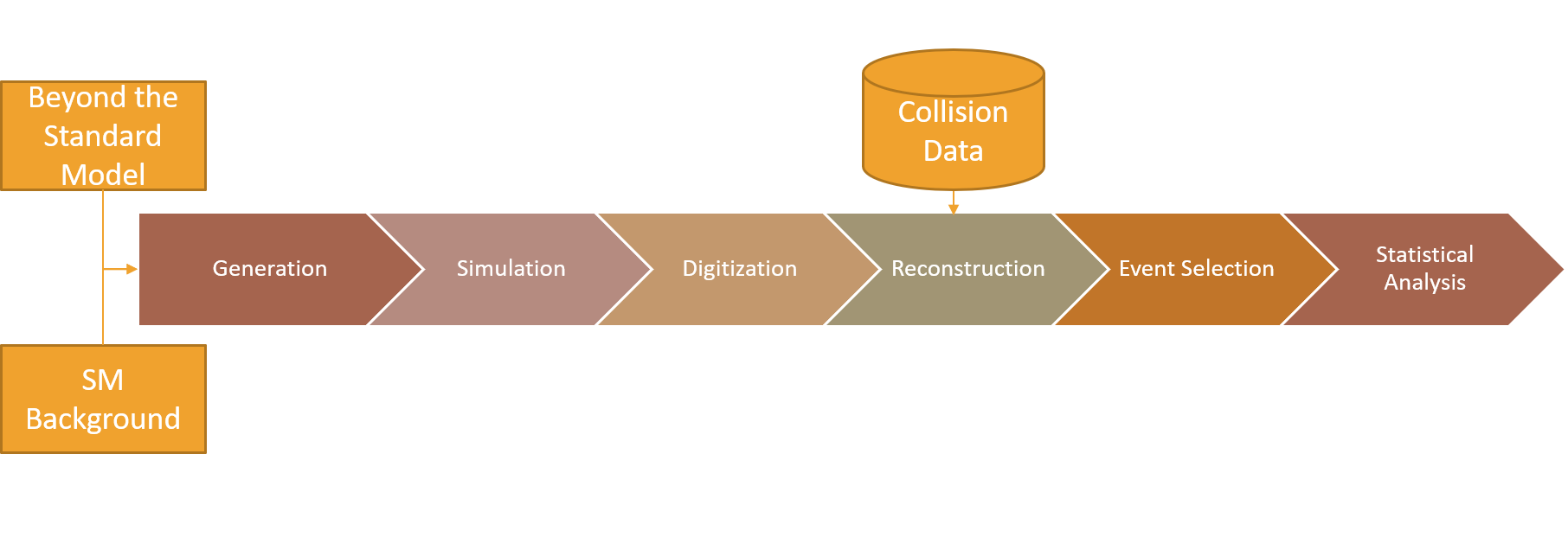}
    \label{fig:search_workflowa}
    }
    \subfigure[]{
    \includegraphics[width=\textwidth]{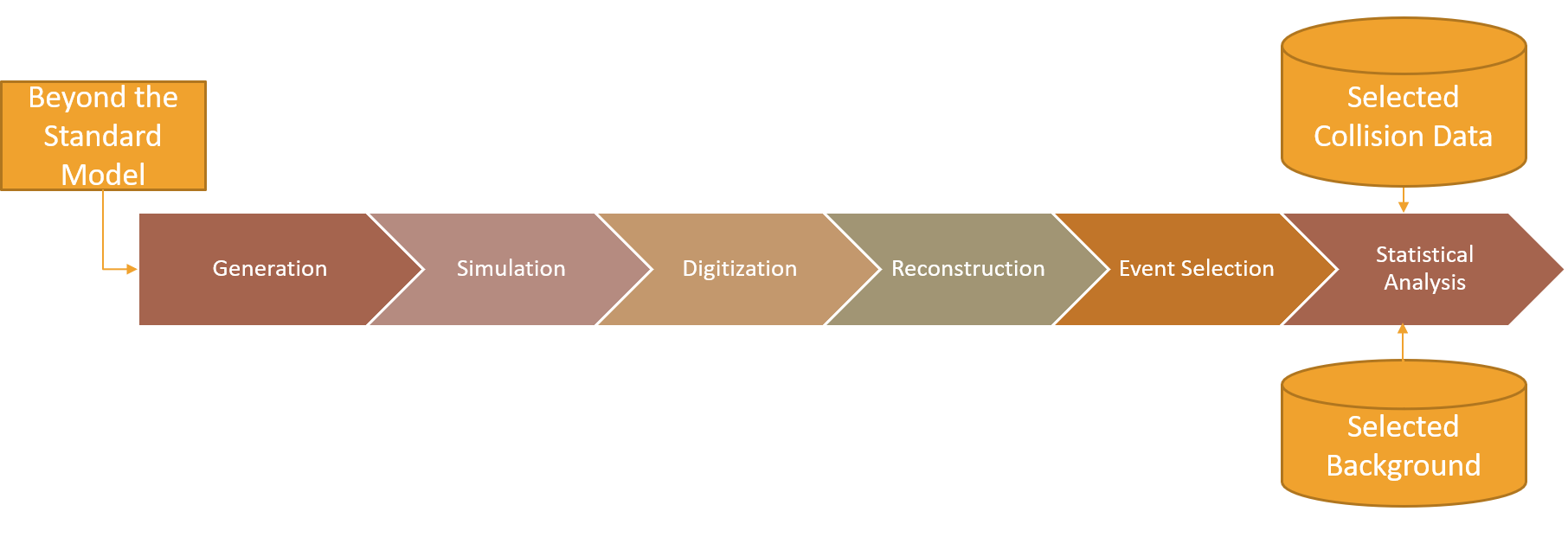}
    \label{fig:recast_workflowb}}
    \caption{An example of a typical search workflow (a) and recast workflow (b) in an ATLAS search for a BSM model.}
\end{figure}

Figure~\ref{fig:search_workflowa} shows a typical search workflow. In order to use the collision data from the detector to discriminate between the BSM and SM theories, we need to know what we should expect to see in each case. Since the details of physics processes with a detector are highly complex, an exact distribution is impossible to predict. Instead, many Monte Carlo simulations are performed to obtain an approximate distribution describing our predictions. The simulation tool-chain is a set of four steps that are highly standardized at ATLAS:

\begin{enumerate}
    \item Generation -- Generation of physics processes using Feynman diagrams based on the theory.
    \item Simulation -- Simulation of the particles as they move through the detector.
    \item Digitization -- Simulation of the digital readings that the detector will record.
    \item Reconstruction -- Reconstruction of the final particles based on the detector's digital readings.
\end{enumerate}

The last step must also be performed on real data, as of course the raw data from the detector is a set of digital readings. Once this is complete, both the simulated data and real data must be passed through a set of event selections, which focus on a particular region of phase space in which we expect there to be large differences between the BSM and SM theories. Finally, the selected data can be passed to statistical software that determines whether there is sufficient evidence for discovery, and if not, which regions of the new model's parameter space can likely be eliminated.

\begin{figure}
    \centering

\end{figure}

\begin{figure}
    \centering
    \includegraphics[width=\textwidth]{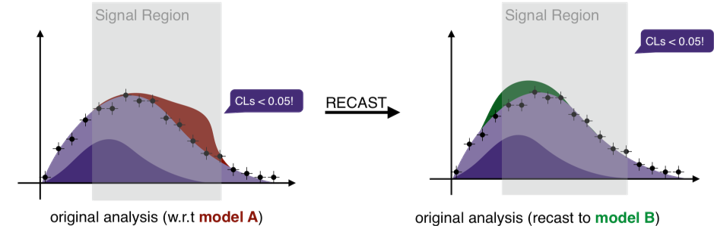}
    \caption{Illustration of the recasting of an analysis originally designed for model A for a new model B. Note that the data (black dots) and background distributions (purple curves) are identical: only the signal distribution changes (red or green curve). See \cite{FNAL}.
}
    \label{fig:recast_idea}
\end{figure}

In contrast, Figure~\ref{fig:recast_workflowb} shows a recast workflow. Note that since we are reusing an existing analysis, we can just use stored collision and background data. All we need to calculate is the prediction of our new signal model. Figure~\ref{fig:recast_idea} shows the distribution of the final observable in the two cases. Note again that the distribution of the background and collision data are equivalent in each case; only the signal changes.

In order to preserve an analysis, three components are necessary:

\begin{enumerate}
    \item Software -- what framework(s) does the analysis use and what are the dependencies?
    \item Commands -- what needs to be done to use the framework(s) for each stage of the analysis?
    \item Workflow -- how are the analysis stages connected?
\end{enumerate}

The first step is accomplished using Docker images, an industry-standard containerization tool. The latter two steps are specified using a Yadage, a workflow language. More details, examples, and discussion on how to integrate with GitLab CI are available at https://recast-docs.web.cern.ch/. Such analysis preservation was recently made a requirement for ATLAS.

\section{Truth-level Reinterpretation}

While the collaboration-specific reinterpretation is very useful, there are a few reasons for alternative approaches:

\begin{enumerate}
    \item 'Full simulation' (modelling detector effects) is computationally expensive and difficult to use.
    \item To determine which regions of phase space would be interesting for a full reinterpretation.
    \item To explore other quantities, such as signal theory uncertainties.
\end{enumerate}

Thus, there was the idea to perform so-called 'truth-level' reinterpretations using recast. In this case, the analysis is performed without modelling the particle's behavior as it interacts with the detector. In this case, we can eliminate the simulation, digitization and reconstruction steps, leaving only three steps: generation, selection and statistics.

\begin{figure}
    \centering
    \includegraphics[width=\textwidth]{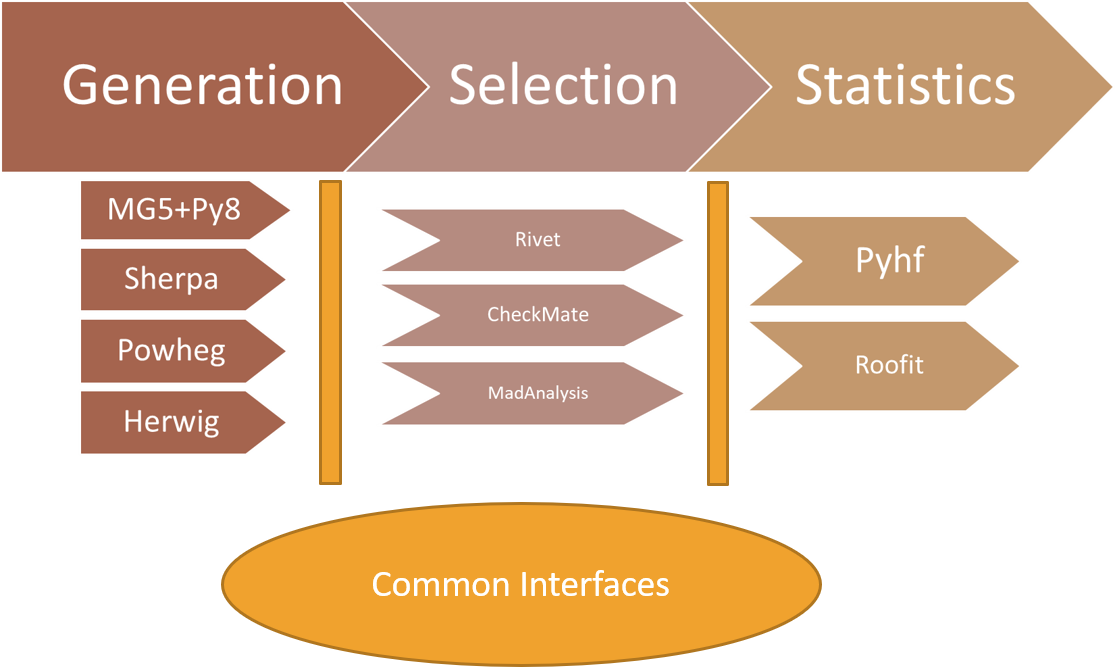}
    \caption{A cartoon illustration of the recast catalogue. For each of the three steps, there are many collaboration-external tools that have been developed. Thus, it is necessary to define a catalogue that describes how the various tools can be put together in a standardized fashion.}
    \label{fig:recast_catalogue}
\end{figure}

Figure~\ref{fig:recast_catalogue} illustrates these three steps. Over the years, many tools have been developed that implement one or several of the steps in different ways. Therefore, we are now developing the 'recast catalogue' that implements standardized sub-workflows for each of these steps, as well as describing how to combine them using common interfaces into a single workflow. The end goal is to facilitate easier truth-level reinterpretation and unify different types of reinterpretation under the same framework. This should also allow for much easier comparison of results across different tools, which can serve as a valuable consistency check, particularly given the complexity of the operations that these tools perform.

\clearpage

\section{Future work}
There are three ideas that we would like to incorporate into the recast framework moving forward:

\begin{enumerate}
    \item REANA backend integration
    \item Web interface
    \item Smart grid selection
\end{enumerate}

Our first goal is integration with the reana backend. Once we have determined an appropriate workflow for either a full or fast reinterpretation, we need to evaluate the workflow and retrieve results. Reana is a cloud-backed computational workflow platform that supports submission of non-interactive batch workflows written in a workflow language such as Yadage. As such, it's a perfect 'backend' computational tool for recast that can enable us to provide faster and more accurate results in greater quantity.

Once we integrate reana, we'd also like to create a unifying web interface for all recast work. Presently, recasting is accomplished through command-line tools, but as part of our goal of making results more accessible to non-experts such as theorists, we want to create a user-friendly web interface. A prototype was created in the past, but work needs to be done to update it to modern standards.

Finally, we'd like to add 'smart grid selection'. Each BSM model typically has some set of parameters that must be specified in order to fully specify the predictions of the model. When performing a search, we evaluate the model on some set of points in that parameter space, and then extrapolate between those points. The set of points we evaluate is known as the 'grid'. The typical approach is to use a uniform grid over a region of parameter space that is thought to be of interest. However, this often involves a lot of wasted computation, as many of the points are far from the contour that describes the bounds on our model. 

\begin{figure}
    \centering
    \includegraphics[width=\textwidth]{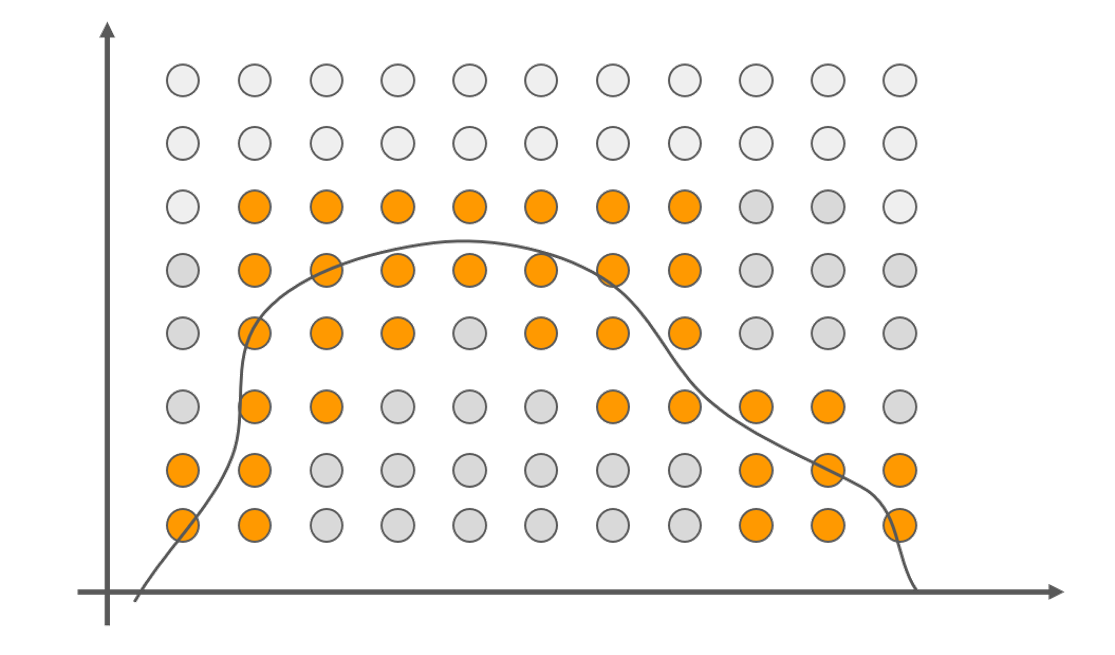}
    \caption{A cartoon illustrating smart grid selection~\cite{excursion}. In a typical uniform grid selection, points far from the contour (gray) are uninformative and result in wasted computation. With smart selection, the search tends to focus on points near the contour (orange).}
    \label{fig:smart_grid}
\end{figure}

Other approaches besides uniform sampling have been developed. In particular, the idea is to implement an active-learning alternative to uniform grid selection that selects points in an iterative fashion, using each evaluation to inform our Bayesian prior~\cite{excursion}. See Figure~\ref{fig:smart_grid} for an illustration.

\begin{figure}
    \centering
    \includegraphics[width=\textwidth]{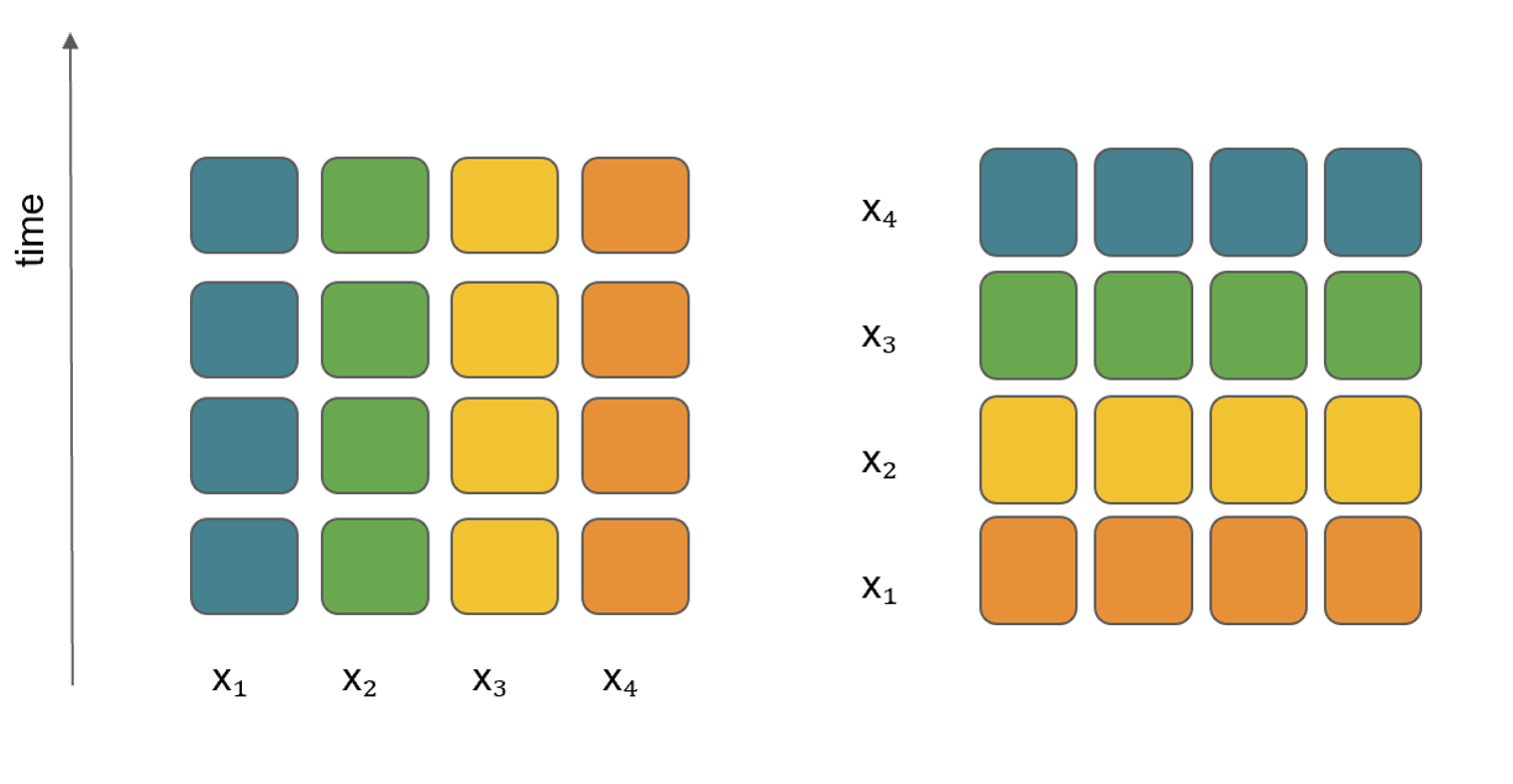}
    \caption{A cartoon illustrating the distribution of finite computing resources for the evaluation of points in uniform grid selection (left) and smart grid selection (right)~\cite{excursion}. Each set of colored blocks corresponds to a single point. As can be seen, in the smart case, even though each point must be calculated sequentially, high utilization is still achieved by parallelizing the computation for each point.}
    \label{fig:wall_clock}
\end{figure}

Typically, such a sequential approach would suffer from being much slower than a uniform grid, where you can exploit parallelization by calculating all points simultaneously. However, the computation of each point is already highly parallelizable. In practice, we only have finite computing resources, so we therefore expect that by shifting parallelization from inter- to intra-point computations, we can reduce both total computation and wall-clock time, as shown in Figure~\ref{fig:wall_clock}.

\clearpage

\section{Conclusion}

RECAST is an analysis preservation and reinterpretation tool that has been used for full reinterpretations that account for detector specifics. Now, recast is being extended to fast reinterpretations that utilize truth-level information that is detector-agnostic through development of a catalogue. This should be useful both for theorists interested in quickly putting limits on their model and for more sophisticated users and developers who are interested in comparing the consistency of the various tools, which should ultimately spur further development. In the future, we have several other ideas planned, including reana integration, a web interface, and smart grid selection and we hope that further contributions are forthcoming.

\section*{Acknowledgements}

\end{document}